 \documentclass[12pt,preprint]{aastex}

 \usepackage{epsfig}

\newcommand{\rs}{r_{_{\rm S}}}

\newcommand{\gapprox}{\lower.4ex\hbox{$\;\buildrel >\over{\scriptstyle\sim}\;$}}
\newcommand{\lapprox}{\lower.4ex\hbox{$\;\buildrel <\over{\scriptstyle\sim}\;$}}
\newcommand{\begeq}{\begin{equation}}
\newcommand{\fineq}{\end{equation}}
\newcommand{\begeqarray}{\begin{eqnarray}}
\newcommand{\fineqarray}{\end{eqnarray}}
\newcommand{\green}{f_{_{\rm G}}}
\newcommand{\Qjet}{L_{\rm jet}}

\newcommand{\SgrA}{Sgr~A$^*$}

\def\ellprime0{\ell'_0}

\shorttitle{Shocks in ADAF Disks}

\shortauthors{Le \& Becker}

\begin{document}

\title{A Self-Consistent Model for the Formation of
Relativistic Outflows in Advection-Dominated Accretion
Disks with Shocks}

\author{Truong Le\altaffilmark{1} and
Peter A. Becker\altaffilmark{2}$^,$\altaffilmark{3}}

\affil{Center for Earth Observing and Space Research, George Mason
University, Fairfax, VA 22030-4444, USA}

\altaffiltext{1}{tler@gmu.edu}
\altaffiltext{2}{pbecker@gmu.edu}
\altaffiltext{3}{also Department of Physics and Astronomy,
George Mason University, Fairfax, VA 22030-4444, USA}

\begin{abstract}
In this Letter, we suggest that the relativistic protons powering the
outflows emanating from radio-loud systems containing black holes are
accelerated at standing, centrifugally-supported shocks in hot,
advection-dominated accretion disks. Such disks are ideal sites for
first-order Fermi acceleration at shocks because the gas is tenuous, and
consequently the mean free path for particle-particle collisions
generally exceeds the thickness of the disk. The accelerated particles
are therefore able to avoid thermalization, and as a result a small
fraction of them achieve very high energies and escape from the disk. In
our approach the hydrodynamics and the particle acceleration are coupled
and the solutions are obtained self-consistently based on a rigorous
mathematical treatment. The theoretical analysis of the particle
transport parallels the early studies of cosmic-ray acceleration in
supernova shock waves. We find that particle acceleration in the
vicinity of the shock can extract enough energy to power a relativistic
jet. Using physical parameters appropriate for M87 and \SgrA, we confirm
that the jet kinetic luminosities predicted by the theory agree with the
observational estimates.

\end{abstract}

% The different journals have different requirements for keywords.  The
% keywords.apj file, found on aas.org in the pubs/aastex-misc directory,
% contains a list of keywords used with the ApJ and Letters.  These are
% usually assigned by the editor, but authors may include them in their
% manuscripts if they wish.


\section{INTRODUCTION}

Radio-loud active galactic nuclei (AGNs) are thought to contain supermassive
central black holes surrounded by hot, two-temperature,
advection-dominated accretion flows (ADAFs) with significantly
sub-Eddington accretion rates. In these disks, the ion temperature $T_i
\sim 10^{12}\,$K greatly exceeds the electron temperature $T_e \sim
10^{10}\,$K (e.g., Narayan, Kato, \& Honma 1997; Becker \& Le 2003). The
observed correlation between high radio luminosities and the presence of
outflows suggests that the hot ADAF disks are efficient sites for the
acceleration of the relativistic particles powering the jets. This
motivated Blandford \& Begelman (1999) to consider the possibility of
strong outflows of matter and energy from such systems, under the
assumption of self-similarity and Newtonian gravity. The same scenario
was extended to the general relativistic case by Becker, Subramanian, \&
Kazanas (2001). However, a single, global, self-consistent model for the
structure of the disk/jet that includes a specific microphysical
acceleration mechanism has not yet appeared in the literature. In
this Letter, we explore the consequences of the presence of a
shock in an ADAF disk for the acceleration of the observed nonthermal jet
particles.

It is well known that for certain sets of outer boundary conditions
describing the specific angular momentum and the specific energy of the
gas supplied to the accretion flow, both shocked and shock-free
solutions for the global disk structure are possible, although there is
an upper limit on the amount of viscosity beyond which no shocks can
form. However, it is clear that shocks can exist in both dissipative and
inviscid flows (e.g., Chakrabarti \& Das 2004). When both shocked and
shock-free solutions are possible for the same set of upstream
conditions, one must incorporate additional physical principles in order
to determine which flow solution actually occurs. Since in general the
shock solution possesses a higher entropy content than the shock-free
solution, we argue based on the second law of thermodynamics that shocks
are preferred (see Becker \& Kazanas 2001). Particle acceleration has
not been studied in shocked disks in either the viscid or the inviscid
cases, and therefore we concentrate here on the simpler situation of
inviscid (adiabatic) flow. In this Letter we focus on disks containing
isothermal shocks because the jump in the energy flux at such a shock
can be identified with the energy carried away from the disk by the
relativistic particles in the outflow.

\section{DYNAMICAL MODEL}

The dynamical structure of the accretion flows considered here follows
the model studied by Chakrabarti (1989a) and Abramowicz \& Chakrabarti
(1990), which comprises an inviscid disk that either contains an
isothermal shock, or else is shock-free. The model is based on the
standard one-dimensional conservation equations for advection-dominated
accretion flows, including utilization of the pseudo-Newtonian
gravitational potential to simulate the effects of general relativity
(Paczy\'nski \& Wiita 1980). In general, the flow is assumed to be
adiabatic and in vertical hydrostatic equilibrium, although the entropy
can jump at the shock. Under the inviscid flow assumption, the
disk/shock model depends on two parameters, namely the energy transport
rate per unit mass, $\epsilon$, and the specific angular momentum,
$\ell$. The value of $\epsilon$ is constant in ADAF type flows since
there are no radiative losses, although it will jump at the location of
the isothermal shock if there is one in the disk. On the other hand, the
value of $\ell$ remains constant since the disk is inviscid. The energy
transport rate per unit mass can be written as
\begeq
\epsilon = {1 \over 2} \, {\ell^2 \over r^2} + {1 \over 2} \, v^2
+ {1 \over \gamma-1} \, a^2 - {GM \over r-\rs} \ ,
\label{eq1}
\fineq
where $v$ denotes the radial velocity (defined to be positive for
inflow), $\rs=2GM/c^2$, and the adiabatic sound speed is given by $a =
(\gamma P/\rho) ^{1/2}$ for a gas with pressure $P$, mass density
$\rho$, and specific heat ratio $\gamma$. Note that $\epsilon$ is
defined to be positive for an inward flow of energy into the black hole.
Furthermore, in this semi-classical approach, $\epsilon$ does not
include the rest-mass contribution to the energy.

Dissipation and radiative losses are unimportant in inviscid,
advection-dominated disks, and therefore the flow is in general
adiabatic and isentropic, except at the shock location. In this
situation, the ``entropy parameter,''
\begeq
K \equiv v \, a^{(\gamma+1)(\gamma-1)} \, r^{3/2} \, (r-\rs) \ ,
\label{eq2}
\fineq
is conserved throughout the flow except at the location of a shock
(Becker \& Le 2003). When a shock is present, we shall use the
subscripts ``-'' and ``+'' to refer to quantities measured just upstream
and just downstream from the shock, respectively. In the isothermal
shock model, $K$ and $\epsilon$ have smaller values in the post-shock
region ($K_{+}$, $\epsilon_{+}$) compared with the pre-shock region
($K_{-}$, $\epsilon_{-}$). The entropy and energy lost from the disk are
carried off by the outflowing high-energy particles, and the total
entropy of the combined (disk/outflow) system is higher when a shock is
present, as expected according to the second law of thermodynamics. By
combining equations~(\ref{eq1}) and (\ref{eq2}) and differentiating with
respect to radius, we can obtain a ``wind equation'' of the form $(1/v)
\, (dv/dr) = N / D$, where
\begeq
N \equiv - \, {\ell^2 \over r^3} + {GM \over (r-\rs)^2}
+ {a^2 \, (3 \rs - 5 r) \over (\gamma+1) \, r \, (r-\rs)} \ , \ \ \ \ \
D \equiv {2 \, a^2 \over \gamma+1} - v^2
\ .
\label{eq2b}
\fineq
Critical points occur where the numerator $N$ and the denominator $D$
vanish simultaneously (see Figure~1{\it b} for a specific example). The
associated critical conditions can be combined with the energy
equation~(\ref{eq1}) to express the critical velocity $v_c$, the
critical sound speed $a_c$, and the critical radius $r_c$ as functions
of $\epsilon$ and $\ell$. The value of $K$ is then obtained by
substituting $r_c$, $v_c$, and $a_c$ into equation~(\ref{eq2}). In
general, one obtains four solutions for the critical radius, denoted by
$r_{c1}$, $r_{c2}$, $r_{c3}$, and $r_{c4}$. Only $r_{c1}$ and $r_{c3}$
are physically acceptable sonic points, although the types of accretion
flows that can pass through them are different. Specifically, $r_{c3}$
is an X-type critical point, and therefore a smooth, global, shock-free
solution always exists in which the flow is transonic at $r_{c3}$ and
then remains supersonic all the way to the event horizon. On the other
hand, $r_{c1}$ is an $\alpha$-type critical point, and therefore any
accretion flow originating at a large distance that passes through this
point must display a shock transition below $r_{c1}$ in order to cross
the event horizon (Abramowicz \& Chakrabarti 1990).

The radius of the isothermal shock, denoted by $r_*$, is determined
self-consistently along with the structure of the disk by satisfying the
velocity and energy jump conditions (Chakrabarti 1989a)
\begeq
R_*^{-1} \equiv {v_+ \over v_-} = {1 \over \gamma \, {\cal M}_-^2}
\ , \ \ \ \ \
\Delta\epsilon \equiv \epsilon_+-\epsilon_- = {v_+^2 - v_-^2 \over 2}
\ ,
\label{eq2d}
\fineq
where ${\cal M}_- \equiv v_-/a_-$ is the upstream Mach number at the
shock location and $R_*$ is the shock compression ratio. Due to the
escape of energy at the isothermal shock, the energy transport rate
$\epsilon$ drops from the upstream value $\epsilon_-$ to the downstream
value $\epsilon_+$, and consequently $\Delta \epsilon < 0$. The flow
must therefore pass through a new inner critical point located at $\hat
r_{c3} < r_*$, which is computed using the downstream value
$\epsilon_+$. This structure represents the transonic solution of
interest here. The kinetic power lost from the disk at the isothermal
shock location is related to observable parameters by writing
\begeq
\Qjet \equiv \dot M c^2 \triangle \epsilon \ ,
\label{eq3}
\fineq
where $\Qjet$ and $\dot M$ are the observed kinetic power and the
accretion rate for a specific source, respectively. Hence we conclude
that $\triangle \epsilon = \Qjet/(\dot M c^2)$. For given observational
values of $M$, $\dot M$, and $\Qjet$, only certain combinations of the
specific angular momentum $\ell$ and the upstream specific energy
transport rate $\epsilon_{-}$ will yield the correct value for
$\triangle \epsilon$. Hence we can view $\epsilon_{-}$ as the
fundamental free parameter and determine $\ell$ using the energy
constraint $\triangle \epsilon=\Qjet/(\dot M c^2)$ once $M$, $\dot M$,
and $\Qjet$ are specified for a particular source. The two sources of
interest here are M87 and \SgrA, which correspond to models~A and B,
respectively.

\begin{figure}
\begin{center}
\epsfig{file=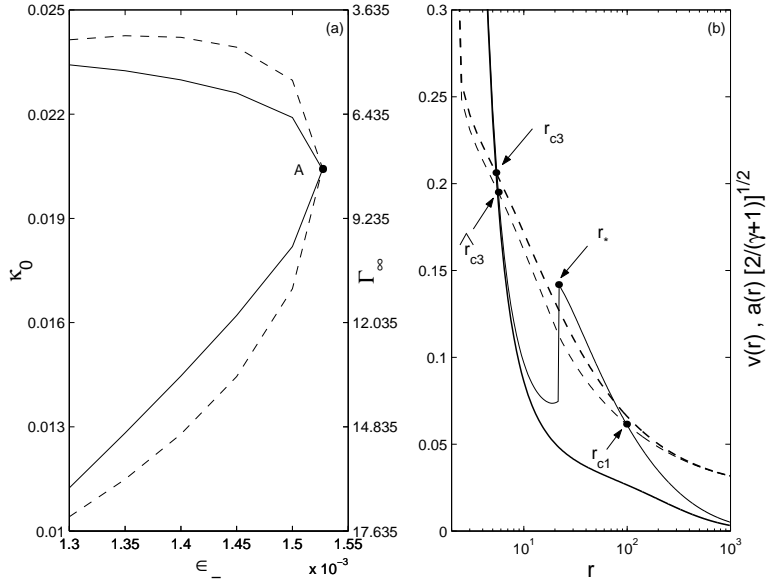,height=8.0cm}
\end{center}
\caption{Results obtained using the M87 values for $M$, $\dot M$, and
$\Qjet$. In panel ({\it a}) we plot $\kappa_0$ ({\it dashed line}) and
$\Gamma_\infty$ ({\it solid line}) as functions of the upstream energy
transport rate $\epsilon_-$ such that the energy conservation condition
$\dot N_{\rm esc} E_{\rm esc} = \Qjet$ is satisfied. In panel ({\it b})
the flow velocity $v(r)$ ({\it solids lines}) and sound speed $a(r)
[2/(\gamma+1)]^{1/2}$ ({\it dashed lines}) are plotted as functions of
the radius for the model~A parameters, which are indicated by the
letter ``A'' in panel ({\it a}). In panel ({\it b}), the thin lines
represent the shocked solutions and the heavy lines denote the
shock-free solutions. See the discussion in the text.}
\end{figure}

\section{PARTICLE TRANSPORT EQUATION}

The model for the diffusive particle transport employed here is similar
to the scenario for the acceleration of cosmic rays first suggested by
Blandford \& Ostriker (1978), although in our situation the shock
maintains a fixed location in the disk and the seed particles for the
acceleration process are provided by the accretion flow itself. The
energy of the injected seed particles is given by $E_0 = 0.002\,$erg,
which corresponds to an injected Lorentz factor $\Gamma_0 \equiv
E_0/(m_p \, c^2) \sim 1.3$, where $m_p$ is the proton mass. Starting
with values for $\epsilon_-$ and $\ell$ associated with a particular
source, we can compute the injection rate for the seed particles, $\dot
N_0$, by setting $\dot N_0 E_0 = \dot M c^2 \triangle \epsilon = \Qjet$
(see eq.~[\ref{eq3}]), which ensures that the energy injection rate for
the relativistic seed particles is equal to the energy loss rate for the
background gas at the isothermal shock location. The particle injection
rate is therefore given by $\dot N_0 = \Qjet / E_0$, which represents
one of the fundamental self-consistency requirements for the model.

The particle Green's function $\green(\epsilon,r)$ describes the energy
and spatial distribution of the relativistic particles injected with energy
$E_0$ at the shock location $r_*$. The associated number and energy densities
of the relativistic particles are given by $n_r=\int_0^\infty 4 \pi E^2 \,
\green \, dE$ and $U_r=\int_0^\infty 4 \pi E^3 \, \green \, dE$, respectively.
Including the escape of the relativistic particles at the shock location,
the energy moments $I_n \equiv \int_0^\infty 4 \pi E^n \, \green \, dE$
of the Green's function satisfy the one-dimensional steady-state transport
equation (Le \& Becker 2004)
\begeq
{d \over dr}\left(4 \pi r H F_n \right)
= 4 \pi r H \left[{(2-n)v\over 3} \, {dI_n \over dr}
+ {\dot N_0 E_0^{n-2} \over 4 \pi r_* H_*} \, \delta(r-r_*)
-c A_0 I_n \, \delta(r-r_*)\right]
\ ,
\label{eq4}
\fineq
where $c$ is the speed of light, $H_*$ is the disk half-thickness
at the shock location, and the flux $F_n$ is defined by
\begeq
F_N \equiv - \, {(n+1)v \over 3} \, I_n - \kappa \, {dI_n \over dr}
\ ,
\label{eq4b}
\fineq
which is obtained by integrating over the vertical structure of the
disk. The terms on the right-hand side of equation~(\ref{eq4}) represent
first-order Fermi acceleration, the particle source, and the escape of
particles from the disk at the shock location, respectively. The terms
on the right-hand side of equation~(\ref{eq4b}) describe the diffusion
and advection of particles, respectively, and $\kappa$ denotes the
spatial diffusion coefficient. The dimensionless parameter $A_0$ is a
constant that represents the microphysical processes controlling the
vertical escape of energetic particles from the disk in the vicinity of
the shock, as discussed below.

To close the system, we must also specify the radial variation of the
spatial diffusion coefficient $\kappa$ in equation~(\ref{eq4b}). Since
the radial dependence of $\kappa$ is not known precisely in this
physical situation, we shall adopt the general form
\begeq
\kappa(r) = \kappa_0 \, v(r) \, \rs \left({r \over \rs} - 1\right)^\alpha
\ ,
\label{eq5}
\fineq
where $\kappa_0$ is a dimensionless positive constant. Close to the
event horizon, inward advection at the speed of light must dominate over
outward diffusion. Conversely, in the outer region, we expect that
diffusion will dominate over advection as $r \to \infty$. A careful
analysis of the differential equation for the relativistic particle
number density $n_r$ obtained by setting $n=2$ in equation~(\ref{eq4})
reveals that these conditions are both satisfied if $\alpha > 1$, and in
our work we shall set $\alpha=2$.

The spatial diffusion coefficient $\kappa$ is related to the magnetic
mean free path via the usual expression $\kappa = c \, \lambda_{\rm mag}
/ 3$. Analysis of the three-dimensional random walk of the escaping
particles then yields
\begeq
A_0 = \left(3 \, \kappa_* \over c \, H_*\right)^2 \ ,
\label{eq6}
\fineq
where $\kappa_* \equiv (\kappa_- + \kappa_+)/2$. We need to specify the
value of $\kappa_0$ in order to compute $\kappa_-$ and $\kappa_+$ using
equation~(\ref{eq5}). Since $A_0$ is independent of the particle energy,
it follows that all of the relativistic particles have the same escape
probability per unit time in the vicinity of the shock, and therefore
the mean energy of the escaping particles is given by $E_{\rm esc} =
U_*/n_*$, where $n_* \equiv n_r(r_*)$ and $U_* \equiv U_r(r_*)$ denote
the relativistic particle number and energy densities at the shock
location. Note that $E_{\rm esc}$ is proportional to $E_0$ and
independent of $\dot N_0$. To obtain a self-consistent result, the
product of the calculated escape energy $E_{\rm esc}$ and the particle
escape rate $\dot N_{\rm esc}$ at the shock location must be equal to
the total energy lost from the background in the dynamical model, and
therefore we set $\dot N_{\rm esc} E_{\rm esc} = \dot M c^2 \triangle
\epsilon = \Qjet$. This condition is not automatically satisfied in
general and therefore it constrains the free parameters $\epsilon_-$ and
$\kappa_0$ as explained below.

\section{APPLICATIONS TO M87 AND \SgrA}

In our numerical examples, the disk structures for M87 and \SgrA are
determined based on observational estimates for the black hole mass $M$,
the mass accretion rate $\dot M$, and the jet kinetic power $\Qjet$.
From observations of M87, we obtain $M \sim 3 \times 10^9 {\rm \
M_{\odot}}$ (e.g., Ford et al. 1994), $\dot M \sim 1.3 \times 10^{-1}
{\rm \ M_{\odot} \ yr^{-1}}$ (Reynolds et al. 1996), and $\Qjet \sim
10^{43-44}\,{\rm erg \ s^{-1}}$ (Reynolds et al. 1996; Bicknell \&
Begelman 1996; Owen, Eilek, \& Kassim 2000). We shall adopt the average
value $\Qjet$ = $5.5 \times 10^{43}\,{\rm erg \ s^{-1}}$. For \SgrA,
observations indicate that $M \sim 2.6 \times 10^6 {\rm \ M_{\odot}}$
(e.g., Schodel et al. 2002) and $\dot M \sim 8.8 \times 10^{-7} {\rm \
M_{\odot} \ yr^{-1}}$ (e.g., Yuan, Markoff, \& Falcke 2002; Quataert
2003). The kinetic luminosity of the jet in \SgrA is rather uncertain
(see, e.g., Yuan 2000; Yuan et al. 2002). In our work, we will adopt the
value quoted by Falcke \& Biermann (1999), $\Qjet \sim 5 \times 10^{38}
{\rm erg \ s^{-1}}$.

Our goal here is to compute self-consistently the mean energy of the
escaping particles, $E_{\rm esc}$, while matching the observational
values of $M$, $\dot M$, and $\Qjet$ for a specific source. We shall
utilize natural gravitational units, with $GM=c=1$ and $\rs=2$, except
as noted. Following Narayan, Kato, \& Honma (1997), we assume
approximate equipartition between the gas and magnetic pressures, and
therefore we set $\gamma=1.5$. Once the values of $M$, $\dot M$, and
$\Qjet$ have been specified for the source, we select a value for the
fundamental free parameter $\epsilon_{-}$ and then compute $\ell$ based
on the energy conservation condition $\triangle \epsilon=\Qjet/(\dot M
c^2)$ (see \S~2). The velocity profile $v(r)$ is determined by
integrating numerically the wind equation $(1/v) \, (dv/dr) = N / D$,
where $N$ and $D$ are given by equation~(\ref{eq2b}). The associated
solution for the isothermal sound speed $a(r)$ can then be computed
using equation~(\ref{eq2}) based on the fact that $K$ is constant in the
adiabatic flow (except at the shock if there is one).

Once the velocity profile $v(r)$ has been obtained, the transport
equation~(\ref{eq4}) can be solved numerically to determine the number
and energy density distributions for the relativistic particles in the
disk, $n_r(r) = I_2(r)$ and $U_r(r) = I_3(r)$. The energy conservation
requirement $\dot N_{\rm esc} E_{\rm esc} = \Qjet$ then allows us to
solve for the corresponding value of the diffusion coefficient constant
$\kappa_0$. Based on the observational values for $M$, $\dot M$, and
$\Qjet$ associated with M87, in Figure~1{\it a} we plot the results
obtained for $\kappa_0$ as a function of $\epsilon_-$. In general, one
obtains two roots for $\kappa_0$ if $\epsilon_- < \epsilon_{\rm max}$,
where $\epsilon_{\rm max} = 0.001527$ for the M87 parameters and
$\epsilon_{\rm max} = 0.001229$ for the \SgrA parameters. No
self-consistent solutions exist if $\epsilon_- > \epsilon_{\rm max}$,
and a single root for $\kappa_0$ is obtained if $\epsilon_- =
\epsilon_{\rm max}$ (see Fig.~1{\it a}). For illustrative purposes, we
shall develop two detailed models by setting $\epsilon_- = \epsilon_{\rm
max}$ for M87 and \SgrA.

The parameter values for the two models are $\epsilon_- = 0.001527$,
$\ell=3.1340$, $\kappa_0=0.02044$, $\dot N_0 = 2.75 \times 10^{46} \,
{\rm s}^{-1}$, $\kappa_* = 0.427877$, $A_0 = 0.0124$, $n_* = 1.19 \times
10^{44}\, {\rm cm}^{-3}$, $U_* = 1.42 \times 10^{42} \, {\rm erg \
cm}^{-3}$, $\dot N_{\rm esc} = 4.61 \times 10^{45} \, {\rm
s}^{-1}$,$E_{\rm esc} = 0.0119\,$erg, $r_* = 21.654$, $\epsilon_{+} =
-0.005746$, $r_{c1} = 98.524$, $r_{c3} = 5.379$, $\hat r_{c3} = 5.659$,
${\cal M}_- = 1.125$, $R_* = 1.897$, and $H_* = 11.544$ for model~A
(M87). For model~B (\SgrA), we have $\epsilon_- = 0.001229$,
$\ell=3.1524$, $\kappa_0=0.02819$, $\dot N_0 = 2.51 \times 10^{41}\,{\rm
s}^{-1}$, $\kappa_* = 0.321414$, $A_0 = 0.0158$, $n_* = 1.93 \times
10^{39}\, {\rm cm}^{-3}$, $U_* = 2.10 \times 10^{37}\,{\rm erg \
cm}^{-3}$, $\dot N_{\rm esc} = 4.56 \times 10^{40}\,{\rm s}^{-1}$,
$E_{\rm esc} = 0.0109\,$erg, $r_* = 15.583$, $\epsilon_{+} = -0.008749$,
$r_{c1} = 131.874$, $r_{c3} = 5.329$, $\hat r_{c3} = 5.723$, ${\cal M}_-
= 1.146$, $R_* = 1.970$, and $H_* = 7.672$. The corresponding dynamical
solutions for $v(r)$ and $a(r)$ (with and without a shock) are plotted
in Figure~1{\it b} for model~A.

The terminal (asymptotic) Lorentz factor of the jet can be computed
using $\Gamma_\infty = E_{\rm esc}/(m_p c^2)$. In Figure~1{\it a} we
plot $\Gamma_\infty$ as a function of $\epsilon_-$ for the M87
parameters. Since in general we obtain two roots for $\kappa_0$ for each
value of $\epsilon_-$, we also obtain two $\Gamma_\infty$ values, with
the larger one corresponding to the smaller $\kappa_0$ root. For
models~A (M87) and B (\SgrA), we obtain $\Gamma_\infty=7.92$ and
$\Gamma_\infty=7.26$, respectively. The associated values for the mean
energy boost ratio, $E_{\rm esc}/E_0$, are given by $E_{\rm esc}/E_0 =
5.95$ and $E_{\rm esc}/E_0 = 5.45$, respectively. Conversely, we find
that in the shock-free models with the same values for $\epsilon_-$,
$\ell$, and $\kappa_0$, the energy is boosted by only a factor of $\sim
1.4 - 1.5$. This clearly establishes the essential role of the shock in
efficiently accelerating particles up to very high energies, well above
the energy required to escape from the disk. The self-consistency
conditions $\dot{N}_0 E_0 = \Qjet$ and $\dot N_{\rm esc} E_{\rm esc} =
\Qjet$ can be combined to conclude that $\dot N_{\rm esc} = \dot N_0 E_0
/ E_{\rm esc}$, which implies that $\lapprox 20\%$ of the injected
particles escape (vertically) through the surface of the disk. The
remainder of the particles are either advected into the black hole or
else diffuse outward (horizontally) through the disk. Our predictions
for the shock/jet location and the asymptotic Lorentz factor agree
rather well with the observations of M87 reported by Biretta, Junor, \&
Livio (2002) and Biretta, Sparks, \& Macchetto (1999). The shock
location predicted by the model in the case of \SgrA is fairly close to
the value suggested by Yuan (2000). However, no reliable observational
estimate for the jet Lorentz factor in \SgrA is currently available.

\section{CONCLUSIONS}

In this Letter, we have demonstrated for the first time the physical
consequences of a standing shock in an inviscid ADAF accretion disk for
the acceleration of relativistic particles and the production of
outflows. The global model presented here provides a single, coherent
explanation for the disk/outflow structure based on the well-understood
concept of first-order Fermi acceleration in shock waves. The theory is
based on an exact mathematical approach to the solution of the combined
hydrodynamical and particle transport equations. Our work may help to
explain the observational fact that the brightest X-ray AGNs do not possess
strong outflows, whereas the sources with low X-ray luminosities but
high levels of radio emission do. We suggest that the gas in the
luminous X-ray sources is too dense to allow efficient Fermi
acceleration of a relativistic particle population, and therefore in
these systems, the gas simply heats as it crosses the shock. Conversely,
in the tenuous ADAF accretion flows studied here, the relativistic
particles are able to avoid thermalization due to the long collisional
mean free path, resulting in the development of a significant nonthermal
component in the particle distribution which powers the jets and
produces the strong radio emission.

We have presented detailed results that confirm that the general
properties of the jets observed in M87 and \SgrA can be understood
within the context of our disk/shock/outflow model. Although the
specific numerical results discussed here are limited to the case of an
inviscid disk, we believe that the inclusion of viscosity will not
change the basic features of the inflow/outflow solutions obtained here
because efficient first-order Fermi acceleration will also occur in
viscous disks, provided they contain shocks. In addition to viscosity,
future work must also address several other important considerations
such as the effects of general relativity and radiative transport on the
disk/shock structure.

The authors are grateful to the anonymous referee for several
useful suggestions that improved the manuscript.


\begin{thebibliography}{}

\bibitem{1} Abramowicz, M. A. \& Chakrabarti, S. K. 1990, \apj, 350, 281

\bibitem{2} Becker, P. A., Kazanas, D. 2001, \apj, 546, 429

\bibitem{3} Becker, P. A., Le, T. 2003, \apj, 588, 408

\bibitem{4} Becker, P. A., Subramanian, P., \& Kazanas, D. 2001, \apj,
552, 209

\bibitem{5} Bicknell, G. V., \& Begelman, M. C. 1996, \apj, 467, 597

\bibitem{6} Biretta, J. A., Junor, W., \& Livio, M. 2002, NewAR, 46, 239

\bibitem{7} Biretta, J. A., Sparks, W. B., \& Macchetto, F. 1999, \apj,
520, 621

\bibitem{8} Blandford, R.~D., \& Begelman, M.~C. 1999, \mnras, 303, L1

\bibitem{9} Blandford, R. D., Ostriker, J. P. 1978, ApJ, 221, L29

\bibitem{10} Chakrabarti, S. K. 1989, Publ. Astron. Soc. Japan, 41, 1145

\bibitem{11} Chakrabarti, S. K., \& Das, S. 2004, \mnras, 349, 649

\bibitem{12} Falcke, H., \& Biermann, P. L. 1999, A\&A, 342, 49

\bibitem{13} Ford, H. C. et al. 1994, \apj, 435, L27

\bibitem{14} Le, T., \& Becker, P. A. 2004, in preparation

\bibitem{15} Narayan, R., Kato, S., \& Honma, F. 1997, \apj, 476, 49

\bibitem{16} Owen, F. N., Eilek, J. A., \& Kassim, N. E. 2000, \apj,
543, 611

\bibitem{17} Paczy\'nski, B., \& Wiita, P. J. 1980, A\&A, 88, 23

\bibitem{18} Quataert, E. 2003, astro-ph/0304099

\bibitem{19} Reynolds, C. S. et al. 1996, MNRAS, 283, L111

\bibitem{20} Schodel, R. et al. 2002, Nature, 419, 694

\bibitem{21} Yuan, F. 2000, MNRAS, 319, 1178

\bibitem{22} Yuan, F., Markoff, S., \& Falcke, H. 2002, A\&A, 383, 854

\end{thebibliography}
\end{document}